# Shaping the notion of #wellbeing in the therapy culture context: an analysis through Instagram narratives

Ainara Larrondo-Ureta, Simón Peña-Fernández and Jordi Morales-i-Gras
*Department of Journalism, University of the Basque Country, Leioa, Spain*

## Abstract

**Purpose** – The aim of this paper is to (1) identify textual and visual themes and sub-themes associated with the #wellbeing hashtag on Instagram, (2) assess their varying levels of engagement and (3) investigate gender bias present in the analysed visual narratives.

**Design/methodology/approach** – This study employs a range of big data analysis techniques to investigate various dimensions of wellbeing on Instagram. Initially, a sample of 9,844 posts was processed using data mining and text analysis methods to identify and categorise significant hashtags, which facilitated the classification of posts into thematic clusters through hierarchical clustering. Engagement was then assessed using non-parametric statistical tests. Additionally, computer vision models were used to analyse and classify visual narratives, grouping images into communities based on visual similarities. Finally, gender representation in the images was examined using object detection models.

**Findings** – The study reveals that the discourse around #wellbeing on Instagram is predominantly feminised and focuses primarily on mental, psychological and spiritual aspects. Therapeutic and positive psychology narratives are the most prevalent and engaging, while physical activity and nutrition play a relatively secondary role. Two major macro-narratives emerge: firstly, mental health and emotional wellbeing, often featuring hashtags related to therapy, self-discovery, spirituality and motivational quotes; secondly, though less prominently, themes and visual narratives concerning physical activity and healthy habits, emphasising exercise and nutrition.

**Originality/value** – The study identifies, for the first time in the academic literature, key themes and sub-themes in the textual and visual narratives on wellness on the social media (Instagram), offering evidence for its feminisation. These findings contribute to the academic discussion on the implications of the notion of wellbeing on Instagram in relation to "therapy culture". This idea can be applied to other social media, as well as to other areas, such as working life, where diverse individualities and groups also coexist.

## 1. Introduction

The concept of wellbeing has expanded in recent decades, going beyond an understanding linked to physical health (wellness), and coming to focus also on the prevention of illnesses and This study is part of the scientific and academic production of the "Gureiker" Research Group (Basque Government). The authors would like to thank the reviewers and editors for their time and dedication in helping to improve this manuscript with their suggestions.

*Funding:* This study was funded by the Basque University System (Gureiker Research Group, IT1496-22). The authors would like to recognise the reviewers and editors for their time and dedication in helping to improve this manuscript with their suggestions.

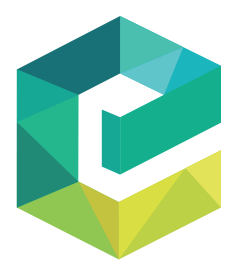

psychological and emotional wellbeing (Kirkland, 2014). In this respect, wellbeing has been characterised as a dynamic notion, in constant creation, which requires consistent attention from individuals, something that is a challenge, particularly for women (Liu *et al.*, 2016).

In this regard, the positive emotion; engagement; relationships; meaning and accomplishment (PERMA) model (Seligman, 2018) is interesting since it favours a multidimensional approach that guides the obtainment of wellbeing based on individual or subjective factors such as positive emotion (P), engagement (E; that is, a psychological connection to activities or organisations), positive relationships (R; integrated socially, cared for, supported by), meaning (M; belief that one's life is valuable, connected to something bigger than oneself) and accomplishment (A; progress towards goals and a feeling of achievement).

The concept of wellbeing has been applied to the digital environment and the social media, where it has been defined as an optimum state of welfare achieved by consumers who consciously manage their use of the social media, aiming to achieve a positive impact on their mental health and keeping in mind their priorities. This means considering the influence of the social media on self-perceptions, personal satisfaction with one's physical appearance, daily activities and, in general, self-confidence. That is to say, those social media with large-scale repercussion on the web, such as Instagram and TikTok, influence the mental wellbeing of people, especially young people.

In this respect, experiences of wellbeing in the context of the social media have been revealed (Liu *et al.*, 2016), taking into consideration that users go to these media to satisfy their psychological and social needs (needs that are cognitive, emotional, personal, related to social integration, etc.) (Lee *et al.*, 2015; Ting *et al.*, 2015). In short, the benefits of consuming subject matter linked to wellness, wellbeing and a good life on the social media have been brought to light, especially during periods when people's physical and mental health is suffering, such as during the Covid-19 pandemic (Biondi *et al.*, 2022). In relation to this utility, it has been demonstrated that a habitual use of social media can negatively influence one's self-esteem and psychological or mental wellbeing (Huang, 2017), although certain studies have indicated slightly more positive conclusions, at least in the specific case of Instagram (Staniewski and Awruk, 2022).

Going beyond considerations of how narratives on the social media affect the physical, mental and emotional wellbeing of those who create them and consume them, it is also worth looking at wellbeing in terms of content, message or narrative in the social media.

While the media and advertising have contributed to increasing individuals' interest in matters linked to wellbeing, especially with regard to physical and mental health (Salmon and Atkin, 2003; Kline, 2006), users themselves have been exploiting, to an ever greater extent, their roles as producers of content, generating, on their own, narratives related to these themes. This has meant that matters such as wellbeing are necessarily covered in a different way to that done in the pre-Internet era, when the mediation of the dominant cultural and content industries was decisive when it came to understanding the concept of wellbeing and related issues. Wellbeing-themed narratives created by individuals on the social media have therefore become a matter of interest, one that has hardly been explored until now.

In this context, it is worth paying attention to the way in which individuals contribute to the definition of wellbeing in the digital conversation in the social media. These media outlets can today be considered one of the most powerful spaces for creation, although at the same time they are characterised by being spaces determined by multiple and complex factors. The social media have the capacity to build overall narratives based on multiple micronarratives, each one of which, on its own, shows a capacity for influence (Staniewski and Awruk, 2022). This makes the subject interesting for fields linked to consumption, marketing, journalism and psychology, among others. Furthermore, the construction of themes, concepts and narratives in the social media is developed based on new frameworks, such as that generated by interactions and engagement with regard to these themes (Rivas-Herrero and Igartua, 2021).

The Instagram service is particularly interesting in this regard, considering its particular communicative aspects, as well as the fact that, at the present time, it is one of the most important scenarios for developing and analysing social interactions. In view of this, this article focusses on the way in which individuals create their narratives on wellbeing and interact directly with wellbeing-related content and with other users via this content.

With this aim, a central part of the study is the analysis of the #wellbeing hashtag. Hashtags and captions facilitate searches for and classification of digital conversations in relation to multiple themes. The also represent an essential component of the image, which is a defining feature of the Instagram platform (Sebastian, 2019).

In view of the absence of research focussing especially on the #wellbeing hashtag in the social media and on Instagram, specifically, this study examines its use aiming to observe its contribution to the conceptualisation of this concept in expansion. This allows a closer look at its characterisation in the specific sphere of the social media, which means not only considering the communicative implications of its use by individuals, but also the way in which individuals themselves contribute to its definition in a context of "therapy culture" (Furedi, 2003). This refers to a context that has grown up in recent decades in which societies show a range of vulnerabilities and promote an "emotional turn" (Furedi, 2003) towards a new culture that frames daily problems from the point of view of the emotions.

Specifically, this study has sought to reveal how the social dialogue on Instagram gives shape to the fashionable notion of wellbeing. To do this, it has considered different research questions (RQ) linked to the #wellbeing hashtag.

Firstly, and starting from the premise that wellbeing is a broad and multifaceted concept, this study asks which are the themes and sub-themes linked to the #wellbeing hashtag, taking into account its combinations with other hashtags. This combination aims to identify thematic similarities or differences, that is to say, whether it is a more or less homogenous or heterogeneous narrative on the concept of wellness on Instagram (RQ1).

Secondly, and considering that some of these most successful narratives in terms of engagement can contribute to generating complex narrative spaces, the study examines the different levels of engagement for these themes or narratives, measurable based on the comments and likes figures (RQ2). Given the specific nature of Instagram, the analysis also asks about other particular questions, such as the visual narratives of the #wellbeing hashtag (RQ3).

In this respect, the analysis identifies groups of images with similar patterns to determine whether the visual element accompanies the heterogeneity predicted in thematic terms in previous research questions. In other words, the study seeks to discover possible associations between the themes and the visual narratives identified about wellbeing. With this purpose, it applies statistical evidence based on association, such as contingency table analysis and the exploration of Cramér's V.

With regard to the visual narratives, the study has also sought to focus attention on the gender representation, and so it looks at the presence of men and women in the images analysed, asking if the visual narrative about wellbeing on Instagram is feminised or masculinised and, in this way, arguing for the presence of some kind of gender predominance (RQ4). In order to do this, computer vision techniques are applied and connections are established between this feminised scenario and the results obtained from the different groups of themes linked to the hashtags.

## 2. Instagram: visual and social hashtagging culture

In a cultural and social setting characterised by the supremacy of the visual and the audio-visual (Hu *et al.*, 2014), Instagram demands particular attention (Leaver *et al.*, 2020). This social platform was launched in October 2010 by Kevin Systrom and Mike Krieger, and since 2012 has belonged to the company Facebook. According to Statista.com (January 2024), in late 2023 this social networking service reached 2.5 billion users, making it the social media outlet with the fourth largest consumption globally. This app is used mainly by young people between 18 and 34, that is to say, it is a social networking service with particular impact in the young adult sector of the population (Huang and Su, 2018).

Instagram represents a social conversation built up based on narratives and interactions promoted through photographs and videos (reels). In fact, it has been understood that Instagram represents the key to "understanding and mapping visual social media cultures" (Leaver *et al.*, 2020, p. 2). In fact, the image makes it possible to attract attention more effectively, as well as promote more concise or specific messages, in this way generating more consistent emotional reactions compared to the written text (Barry, 1997). In short, images exercise a major influence on the development of individuals' identity, self-perception and awareness (Hill, 2004).

The use of this social media service responds to different motivational structures and can even create considerable attachment among users, a circumstance defined with the name "intrusion" (Rivas-Herrero and Igartua, 2021). Excessive participation on Instagram and the degree of dependence has been linked directly to the communicative characteristics of this networking service, which facilitate uses based on brief and fast communicative exchanges via images, although it also allows the use of written language. The interaction can be done in different ways, principally by means of comments and the use of hashtags and direct messages.

Like other social media platforms, Instagram functions in a way determined by a system of algorithms that act to process the information generated by users (likes, shared content, searches, geographical locations, etc.) and offer, as a result, content that potentially responds to their interests and cognitive frameworks (Bucher, 2012). In general, it has been understood that the narratives created based on metacommunicative images and messages offer greater interest and engagement for audiences, something that is perceptible or measurable through the indication of likes and comments (Romney and Johnson, 2018).

Hashtags have been an essential resource on Instagram since the launch of the platform in January 2011. Hashtags can be defined as labels or words preceded by the hashtag symbol, "#", which offer information about the content of the photo. This orders images thematically and enables attention to be captured with respect to themes, and also helps users search for images and increases their visibility. In this way, hashtags generate narratives composed of multiple microstories grouped under a single theme -that of the hashtag-, thus appealing to and involving users, creating a kind of digital community (Giannoulakis and Tsapatsoulis, 2016). In short, hashtags represent "narrative tools", which makes them a decisive element for the circulation of content in the social media, although this question has hardly been looked at in the context of discourse and sociolinguistic analysis (Giaxoglou, 2018).

In this regard, it seems clear that users employ hashtags in a way that goes beyond the functional (classifying, informing, etc.), since hashtags also have a metacommunicative purpose, based on codes such as "emphasising", "iterating", "critiquing", "identifying", and "rallying" (Daer *et al.*, 2014, pp. 12–16). It can also be understood that hashtags involve certain doses of emotionality (Small, 2011; Mohammad and Kiritchenko, 2015), particularly when they are used by women (Giannoulakis and Tsapatsoulis, 2016).

The matter of the use of hashtags in the creation of narratives and stories in the social media has been tackled in various previous studies, in which in the hashtag is characterised as a social and discursive practice with the capacity to influence the narrative positions of online users (Androutsopoulos, 2014; Giaxoglou, 2018). Among other outstanding findings, the possibility has been observed for hashtags to encourage mobilisation and activism towards certain causes represented by hashtags, which has come to be called "hashtivism", widely analysed in the case of #BlackLivesMatter (Edrington, 2022) and #Metoo (Hillstrom, 2018), to cite some representative examples.

Past studies, such as that by Giaxoglou (2018) of Twitter, suggest that the exchange of hashtags can introduce changes in the accustomed parameters for understanding a theme or event and have an even greater influence than the traditional influence of agents that are legitimised in terms of public opinion, such as the media and advertising. This necessarily introduces specific parameters into the understanding of the form in which certain themes which trigger great interest among users are discussed, defined and interiorised (Giaxoglou, 2018).

Although so far there has been no research done on the #wellbeing hashtag, there have been studies of others related to physical and mental wellbeing, such as #goodlife. This study enabled the classification of narratives linked to wellbeing and determined, for these, different levels of environmental and social involvement, as well as popularity, with three standing out: the good life of the self-made affluent entrepreneur, the good life of the world-traveller and the good life as shared experience (Loukianov *et al.*, 2020).

All the aspects and studies cited make clear the multi-faceted nature of narratives on Instagram and the influence that hashtags have on them. In this regard, it would be important to remember that this potential has also been called into question by the capacity of hashtags to reproduce gender biases and other kinds of inequalities (Sebastian, 2019).

## 3. Methodology

In order to look at the concept of wellbeing (RQ1), the study had a sample that consisted of 9,844 posts published on Instagram with the #wellbeing hashtag during the month of October 2023. These data were acquired by means of the Ensembledata supplier and were processed using various analytical operations, given below.

In the first place, the 500 most important hashtags in the conversation were selected, including #wellbeing itself, and these were used to construct a vector space that would allow the posts to be classified. This operation was carried out with Orange Data Mining for Python 3 (Demsar *et al.*, 2013). Using the "preprocess text" widget, texts were put into lower case, accents were eliminated and the 500 most frequent hashtags in the dataset were selected by means of a tokenisation based on the regular expression "#\w+". Then, the Bag of Words algorithm was applied in order to count the frequency of appearance of each hashtag in each post, generating a vector space of 500 parameters. The vector space defined by the hashtags made it possible to identify groups of posts that have used the same hashtags.

The next step was to calculate the distances between the rows of the dataset (i.e. the posts) using the cosine distance metric, calculated based on the 500 vectors created previously. This metric was selected because of its performance in multi-dimensional contexts (France *et al.*, 2012), offering a distinct advantage in accurately assessing similarities in content orientation without being affected by vector magnitude. Such characteristics are crucial when comparing posts based on their semantic similarity — posts utilising the same hashtags are deemed closer, whereas those with different hashtags are considered more distant. This approach ensures that the analysis focuses more on the content similarity rather than the volume of hashtags used, enhancing the relevance and precision of the findings.

The study's goal has not so much been to identify the specific distance of a post or group of posts in relation to others, but rather to group together posts that have used the same hashtags, in order to identify different uses of them. With this aim, the images were grouped together in clusters using hierarchical clustering (Murtagh and Contreras, 2012). This technique constructs a hierarchy of clusters by progressively merging the closest pairs, resulting in a tree-shaped structure known as a dendrogram. This visual representation is particularly useful for observing groupings of images according to levels of similarity. Hierarchical clustering offers a unique advantage by allowing analysts to explore the data at different levels of aggregation, from fine to coarse groupings, making it possible to discern the subtle nested relationships among the images. This method not only highlights the primary clusters but also how these clusters are interconnected, providing deeper insights into the underlying patterns in the dataset. Choosing a number of clusters in a hierarchical clustering is always a complicated task, given that, although a smaller number of groups means that these groups are more inclusive and less discriminating, this can also mean that their later qualitative analysis will be more complex or difficult to carry out. In the case of this study, the fusion of the clusters was determined with the Ward method, which gradually merges together clusters that involve the minimum increase in the cluster's total variance. A figure of 10 clusters was chosen, after it was seen that a sufficient thematic variation could be observed with these data.

Then, the clusters were described qualitatively by means of the observation of the main hashtags for each of them. More specifically, the 10 most important hashtags for each cluster were selected, and particular attention was paid to the combined use of these, which has given rise to different patterns described in the results section. Furthermore, the images published in the posts with most likes for each cluster were identified to find the most successful visual narratives for each thematic cluster, although not necessarily the most common or the most prototypical narratives.

The combination of the above referred two techniques made it possible to identify different kinds of themes linked to the hashtag #wellbeing, thus answering the first research question.

With regard to the second research question (RQ2), statistical techniques were applied in order to identify the differences of engagement registered in the clusters identified in the previous stage. Since the data did not have conditions of equality of variances and of normality, ANOVA tests were run in a non-parametric way, using the Kruskal–Wallis model with the R-based Jamovi software (Sahin and Aybek, 2019). As well as the statistical significance, attention was paid to the $\varepsilon$ test in order to assess the effect size of the relationship between the grouping variable — the cluster identified using hierarchical clustering — and the dependent variable — the engagement metric selected on each occasion.

Concerning the third research question (RQ3), in order to identify the visual narratives linked to the #wellbeing hashtag, a similar strategy was used for the classification of the posts was followed, although instead of using 500 attributes contained in the data itself (the 500 most relevant hashtags), a pre-trained neural network was used to classify the posts' images.

For this, the images were vectorised using OpenAI's CLIP model [1], which has been freed and made available on the popular model service Transformers Hugging Face. The computer vision model consists of a zero-shot classifier, which means that it is able to detect a large number of different elements in images, allowing users to determine which elements they wish to identify, instead of being set up to identify a finite set of objects. It is precisely the fact that it is a zero-shot model that makes it especially interesting for this study, by enabling the generation of a vectoral space of images linked by patterns of similarity and dissimilarity.

As in the previous case, the cosine distances between the images, based on their vectors, were calculated. On this occasion, however, instead of a distance matrix and hierarchical clustering, it was chosen to create a graph with the images as vertices and the cosine similarities as edges, in order to make better use of the characteristics of the multi-dimensional space sketched out by the 768 parameters of the CLIP model. The graph only included distances between vertices that were equal to or greater than a threshold of 0.7, and the final design consisted of 9,316 vertices and 83,858 edges that captured similarity relationships among the vertices. This represents 94.64% of the posts captured, leaving to one side those that present duplicate images.

In order to identify communities, the Louvain algorithm [2] was used on the graph, grouping those vertices that share features. Using this procedure, it was possible to identify groups of similar images, which constituted visual narrative units in a field of images united by patterns of similarity. In this case, the aim was not so much to identify the most successful narratives in terms of engagement created, but rather the most habitual and common.

The visual groups identified were analysed with a qualitative strategy, that is, scrutiny was focussed on the most numerous groups of the graph —those involving at least 1% of the vertices. Then, the link between the visual narratives identified —the communities or groups of similar images— and the themes identified based on the uses of hashtags — the clusters identified previously — was analysed. This analysis was carried out based on contingency tables and using the Cramér's V statistic, able to measure the intensity of the link between qualitative variables.

Finally, the presence of women and men in the images identified on Instagram on the topic of #wellbeing was measured (RQ4). With this purpose, firstly, the images in which only a single person appears were identified using Meta's DETR model [3], especially trained to identify particular parameters in images, including people and objects. In total, people were identified in 2,199 images. Secondly, OpenAI's CLIP model was applied once again in order to determine whether the person that appears in the image is a woman or a man, using its zero-shot classification capacity.

## 4. Results

### *4.1 Themes linked to #wellbeing (RQ1)*

With the vector space of 500 hashtags defined for each post acquired, the study has made it possible to identify and view 10 clusters. The dendrogram generated by Orange (Figure 1) of

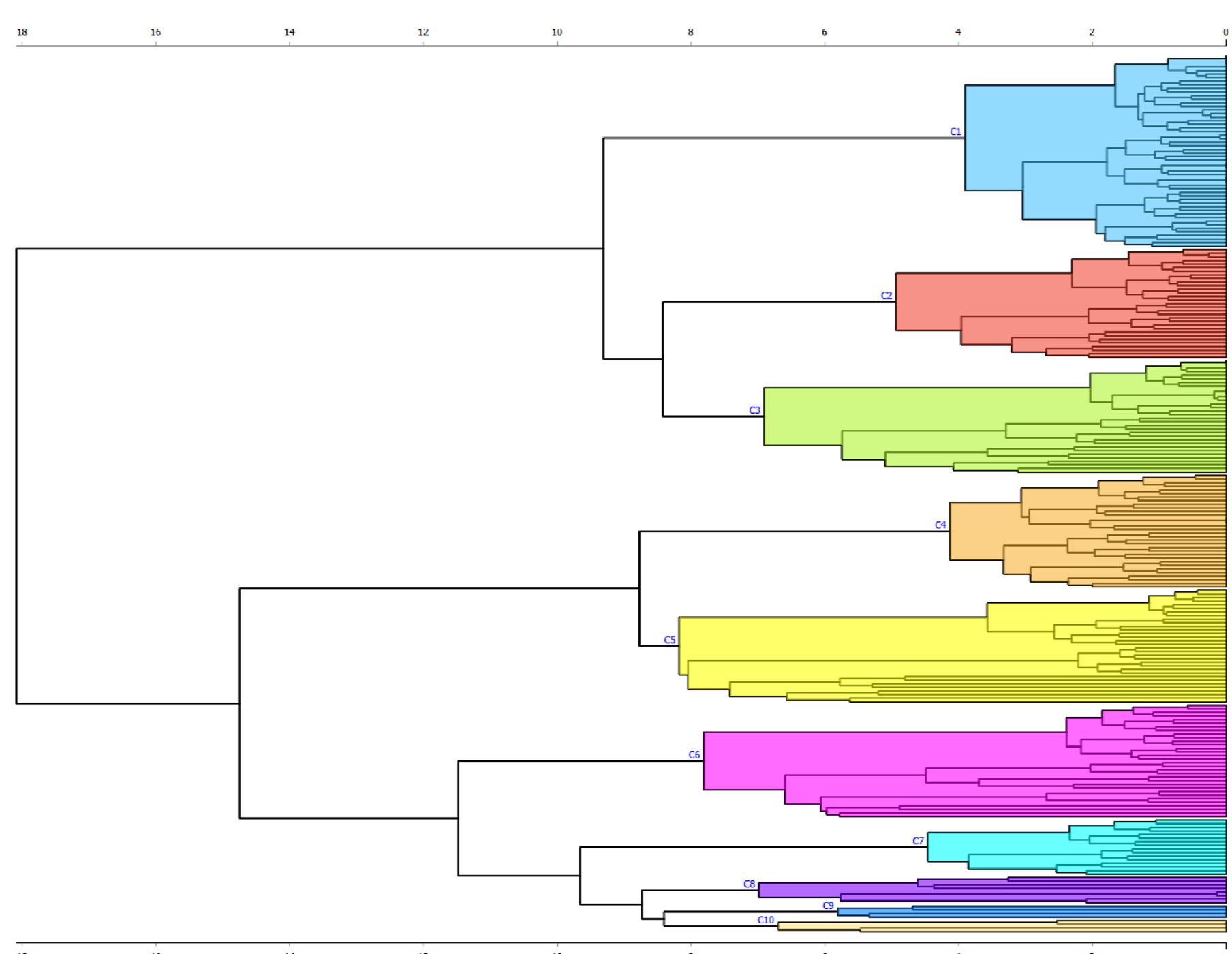

**Figure 1.** Dendrogram created based on the vector space of 500 hashtags

**Source(s):** Own elaboration with Orange Data Mining (Python)

the 10 clusters allows an identification of different groups of themes sufficiently broad and inclusive. This number of clusters has been determined using the cut-off point, which is a common procedure for this unsupervised machine learning analysis technique. In any case, this does not imply that a decision was made regarding where to situate each post, or that a choice was made regarding merging or separating specific clusters.

As shows Figure 1, the clusters identified in the dendrogram pruned to a depth of 8 points, with pruning being an operation that influences how the dendrogram is displayed, but not how the clusters are calculated. The size of our clusters is variable, while some of them do not reach 400 posts (clusters 1, 2, 4, 7 and 10), others gather over 2000 posts (cluster 5), or even over 3000 (cluster 6). This means that not all the themes identified have the same numerical importance: themes 5, 6 and 9 gather practically 70% of the posts in the dataset, while of the others, only number 3 has more than 5% of the posts. A first consideration that it is possible to state from the results of this study is, therefore, that the different narratives identified have levels of presence and are represented to extents that are very unequal, giving rise to majority narratives, on the one hand, and minority narratives, on the other (Table 1).

When users talk about #wellbeing on Instagram, they tend to use more or less the same hashtags: only 51 unique hashtags appear among the 10 most frequent hashtags for the 10 clusters, which is evidence of a shared conversational framework for those who talk about #wellbeing on Instagram (Table 2). This result indicates rather a homogenous narrative. The hashtags that occur most repeatedly are #wellness (in all 10 clusters), #selfcare (in 9 clusters), #mentalhealth and #health (both in 7 clusters) and #mindfulness (in 6 clusters). Many of these recurring hashtags, as we will see, appear linked to the world of positive psychology, coaching, therapy culture and self-help, as well as spiritual matters. The other hashtags that occur repeatedly do so in only 2 or 3 clusters, and a total of 35 appear only in one cluster.

With a finer analysis it is possible to observe how, in the clusters, the emphasis is placed on different matters, such as mental health and emotional wellbeing (clusters 1 to 3), physical activity and healthy habits (clusters 4 and 5), holistic and spiritual wellbeing, personal care and motivation (clusters 6, 7, 8 and 9) and mindfulness and meditation (cluster 10). These are interlinked themes and they could certainly be grouped into larger and more inclusive clusters, such as, on the one hand, physical wellbeing as a result of sport, exercise and good nutrition (clusters 4 and 5), and on the other emotional, mental and spiritual wellbeing resulting from positive psychology, coaching, mindfulness, from beauty rituals or from some kind of activity presented as therapeutic or religious (the other clusters).

The narrative around #wellbeing on Instagram is thus divided into different sub-themes and it puts more emphasis on some matters than on others. In this regard, it is undeniable

| Cluster number | Counts | % of Total | Cumulative % |
|---|---|---|---|
| 1 | 259 | 2.6% | 2.6% |
| 2 | 367 | 3.7% | 6.4% |
| 3 | 822 | 8.4% | 14.7% |
| 4 | 338 | 3.4% | 18.1% |
| 5 | 2013 | 20.4% | 38.6% |
| 6 | 3425 | 34.8% | 73.4% |
| 7 | 315 | 3.2% | 76.6% |
| 8 | 461 | 4.7% | 81.3% |
| 9 | 1391 | 14.1% | 95.4% |
| 10 | 453 | 4.6% | 100.0% |

**Source(s):** Own elaboration con Jamovi (R)

**Table 1.** Frequencies in the clusters: How many posts represent each theme?

**Table 2.** The 10 most frequent hashtags for each cluster

| Cluster 1 | Cluster 2 | Cluster 3 | Cluster 4 | Cluster 5 |
|---|---|---|---|---|
| #mentalhealth | #mentalhealthmatters | #mentalhealth | #fitness | #wellness |
| #wellness | #mentalhealthawareness | #selfcare | #health | #health |
| #health | #mentalhealth | #therapy | #gym | #healthylifestyle |
| #fitness | #selfcare | #selflove | #exercise | #healthyliving |
| #coaching | #mentalhealthsupport | #mindfulness | #wellness | #nutrition |
| #mindfulness | #wellness | #wellness | #workout | #selfcare |
| #selfcare | #therapy | #anxiety | #fitnessmotivation | #fitness |
| #mentalhealthawareness | #mentalwellness | #healing | #motivation | #healthy |
| #mentalwellbeing | #psychology | #mentalhealthawareness | #personaltrainer | #mentalhealth |
| #physicalhealth | #mindfulness | #psychology | #strength | #lifestyle |

| Cluster 6 | Cluster 7 | Cluster 8 | Cluster 9 | Cluster 10 |
|---|---|---|---|---|
| #mentalhealth | #selfcare | #massage | #selfcare | #mindfulness |
| #wellness | #selflove | #selfcare | #selflove | #meditation |
| #community | #wellness | #wellness | #wellness | #yoga |
| #nature | #mindfulness | #relax | #love | #wellness |
| #health | #healing | #beauty | #mindfulness | #health |
| #selfcare | #health | #relaxation | #mentalhealth | #mentalhealth |
| #support | #balance | #serum | #motivation | #healing |
| #leadership | #beauty | #health | #happiness | #love |
| #coaching | #spirituality | #massagetherapy | #inspiration | #retreat |
| #art | #meditation | #spa | #mindset | #selfcare |

**Source(s):** Own elaboration

that the difference of emphasis exists: when more emphasis is placed on mental health, it is not put on the spiritual and the religious, and when it is put on the spiritual and the religious it is not put on massages or beauty therapies and rituals. However, the data also enable to observe rather similar narrative dimensions, which share a central thread rooted in therapy culture, self-care and positive psychology. Therefore, while the #wellbeing narrative field on Instagram is rich in nuances —that are probably very important for users who feed the field with their posts and their activity in the social media—, it is also a relatively homogenous narrative field, or at the very least one that has a strong discursive thread.

Attending to the five images that accompany the most successful posts for each cluster, those clusters which have a more notable orientation towards emotional, mental and spiritual wellbeing (all except 4 and 5) and those more oriented towards sports and nutrition (clusters 4 and 5) have been considered separately.

In general terms, various elements that are common to a large number of clusters are observed, such as inspiring and motivational phrases (clusters 1, 2, 3, 5, 6 and 10), which reinforces the idea described above, referring to the existence of a common narrative around the world of #wellbeing linked to positive psychology, coaching and therapy culture. This kind of content often consists of a simple montage with an inspiring message (e.g. "the calmer you are, the clearer you think", cluster 6), or even a list of precepts or instructions presented simply in order to increase our wellbeing or, more specifically, to overcome a depressed state (e.g. "things to try when you don't feel like getting out of the bed in the morning", cluster 2).

The analysis of the 5 most important images for each cluster also suggests that there are some which are more thematised, for example 4 — oriented strongly towards physical exercise and sport — or even 7 and 8 — oriented towards the world of products and beauty routines). Nonetheless, the majority of the images that appear in the classification could well appear in any cluster, which reinforces the notion that the narrative we are exploring comes across as essentially homogenous.

*4.2 Differences of engagement in the themes identified (RQ2)*

The analyse of the differences in the engagement figures and the posting formats used for each theme identified shows a statistically significant positive correlation between likes and comments in the posts ($R = 0.49$, $p$-value <0.001), indicating that, for more likes, more comments. In terms of engagement figures, however, the moderate intensity of the correlation suggests a certain specialisation in the contents of successes, in the sense that some of them will stand our much more for their likes, while others will do so for their comments.

To identify the differences in the likes and comments figures among clusters we have applied analysis of variance tests. Since the conditions of equality of variances and of normality in the data have not been met, we chose to implement the Kruskal–Wallis test, also known as the non-parametric ANOVA test, which is the appropriate test when the conditions of homogeneity of variances or of normality required by the data in a parametric ANOVA are not met.

Firstly, we have applied the test for the dataset, considering the figure of likes and the figure of comments as the dependent variable and the cluster as the grouping variable. Both associations appear as statistically significant ($p$-value <0.001), but with a negligible effect size ($\varepsilon^2 = 0.013$ for the likes, $\varepsilon^2 = 0.040$ for the comments). This means that, although the theme probably is related to the engagement produced, there are many other factors that influence this, starting with the user who publishes the content and the number of followers and engagement in previous posts — we already know that social media algorithms promote the content of the most successful users, making them even more successful.

For this reason, we have opted to replicate the previous analysis for the 100 most successful contents from each cluster: a dataset was generated with the 100 posts with most likes for each cluster and another with the 100 posts with most comments for each cluster. After applying the non-parametric ANOVA tests for these filtered datasets, statistically significant results in all cases ($p$-value <0.001) and very important effect sizes were obtained. The effect size for the differences in the figures of likes among clusters for the 100 contents with most likes for each cluster is $\varepsilon^2 = 0.46$, and the effect size for the differences among the comments figures among clusters for the 100 contents with most comments for each cluster is $\varepsilon^2 = 0.49$.

As this suggest, the differences among clusters become important and can explain practically 50% of the variability of the dependent variable when the attention is put on outstanding Instagram contents for each of the clusters and on the specific engagement metric. Therefore, even if the general norm is that the #wellbeing hashtag posts are going to achieve few likes and few comments, whatever theme they address, when we focus on the specific segment of the most liked or commented contents, then the differences among clusters become crucial in order to understand this engagement. In other words, when we pay attention to the specific segment of the successful cases, then the thematic differences become important in order to understand the amount of success.

In the case of the contents with the most likes (Figure 2), cluster 6 — oriented to mental health and holistic wellbeing, but also to leadership and coaching — is the one that receives the most likes. This is followed, at some distance, by cluster 9 — oriented to personal care and mindfulness, but also to motivational and inspiring narratives. With regard to the contents with most comments (Figure 3), the same clusters stand out in this respect, followed by 5 and 3, linked to matters of mental health and therapy, in one case, and physical activity and nutrition, in the other.

*4.3 Exploration of visual narratives (RQ3)*

After creating the graph with images linked based on their level of similarity, according to the vectors generated by implementing the CLIP model to the set of non-duplicated images, groups or communities of similar images were identified using the Louvain algorithm. In total, 4,906 communities were identified with a Modularity of 0.527. Note that these communities are different from those previously detected by hierarchical clustering. Previously hashtags were used and now are embeddings; previously hierarchical clustering

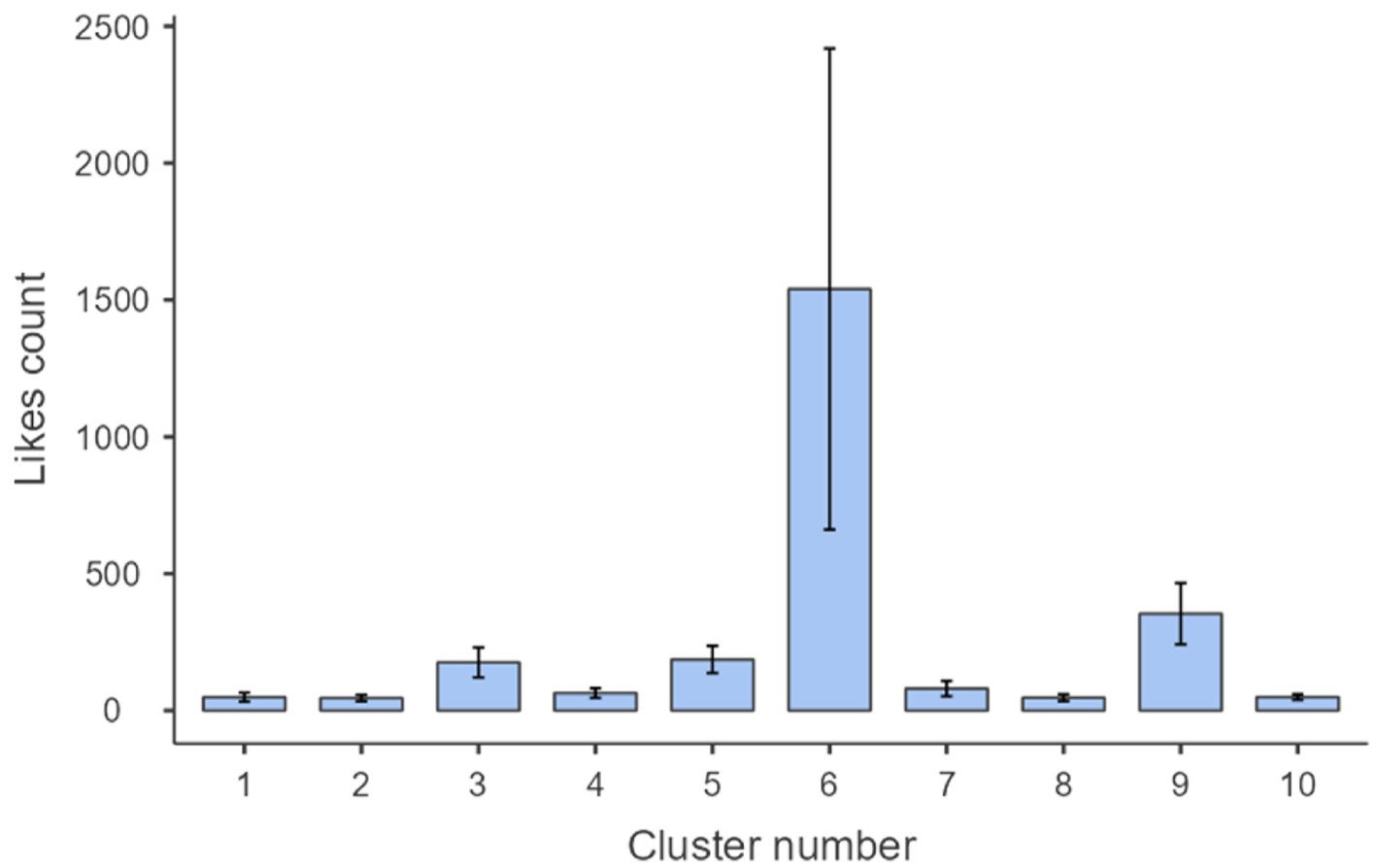

**Source(s):** Own elaboration with Jamovi (R)

**Figure 2.**
Average likes per post in the 100 contents with most likes in each cluster

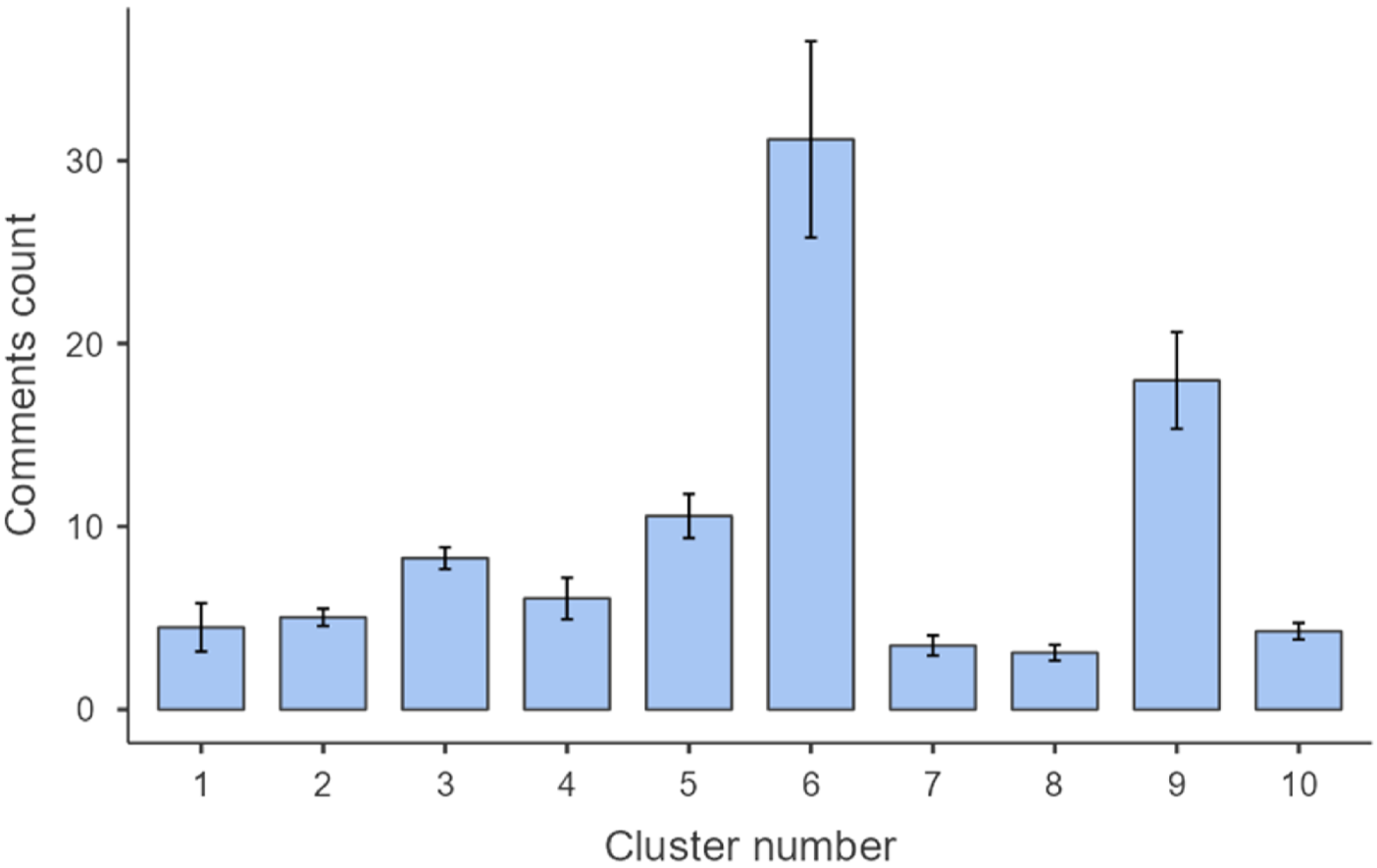

**Source(s):** Own elaboration with Jamovi (R)

**Figure 3.**
Average comments per post in the 100 contents with most comments in each cluster was used and now the Louvain algorithm is used. Later, in the next section, we will explore the intersections between clusters (i.e. emerging from hashtag analysis) and communities (i.e. emerging from image analysis).This suggests a strongly segmented scenario with visual narratives heterogeneous and varied, in which the vertices appear very strongly linked with those of the same class and very weakly with others. Only 7 groups manage to gather over 1% of the vertices and these, in total, represent 39.22% of the vertices, but gather 96.9% of the graph's edges. This means that 60% of the network's vertices form part of narratives that gather less than 1% of the images: small, isolated communities that, as a rule, are differentiated from the others from the viewpoint of their visual narrative. The graph above shows the 7 main groups or narratives, according to the number of images that represent them, accompanied by some images that are very prototypical in the cluster.

The images of the 7 communities accumulate a high degree of weighed input, which reflects a high volume of links with the other images in the graph. These visual narratives relate to elements such as inspiring quotes —generally cards with simple montages—, images of natural and outdoor scenes that evoke peace and tranquility, images of bodily massages and therapies, images that show people relaxing, images of yoga and Pilates, images of selfies of women and images of visually striking healthy foods. It is worth highlighting the fact that the group of images on the subject of relaxation (community 43) represents the cluster's most central narrative, found, as it is, in the middle of the others and acting as a central thread among narratives.

Regarding specifically the association between themes and visual narratives, results show that the 4,906 visual narratives are associated, in a statistically significant and substantial manner —*p*-value <0.001 and the Cramér's V of 0.21 — with the themes or thematic clusters identified in the analysis of the textual contents or hashtags of the posts. This can be interpreted as an effect somewhere between moderate and high.

Figure 4 shows the difference between the frequencies observed and expected for each intersection among the clusters of themes identified using the hashtags — the vertical axis — and the 7 most important visual narratives detected in the images graph —the horizontal axis—. The outstanding findings include the positive associations between clusters 2 and 3 (mental health and psychological therapies) and the visual narrative 6 (inspiring quotes), between cluster 5 (healthy nutrition) and the visual narratives 43 and 88 (relaxation and

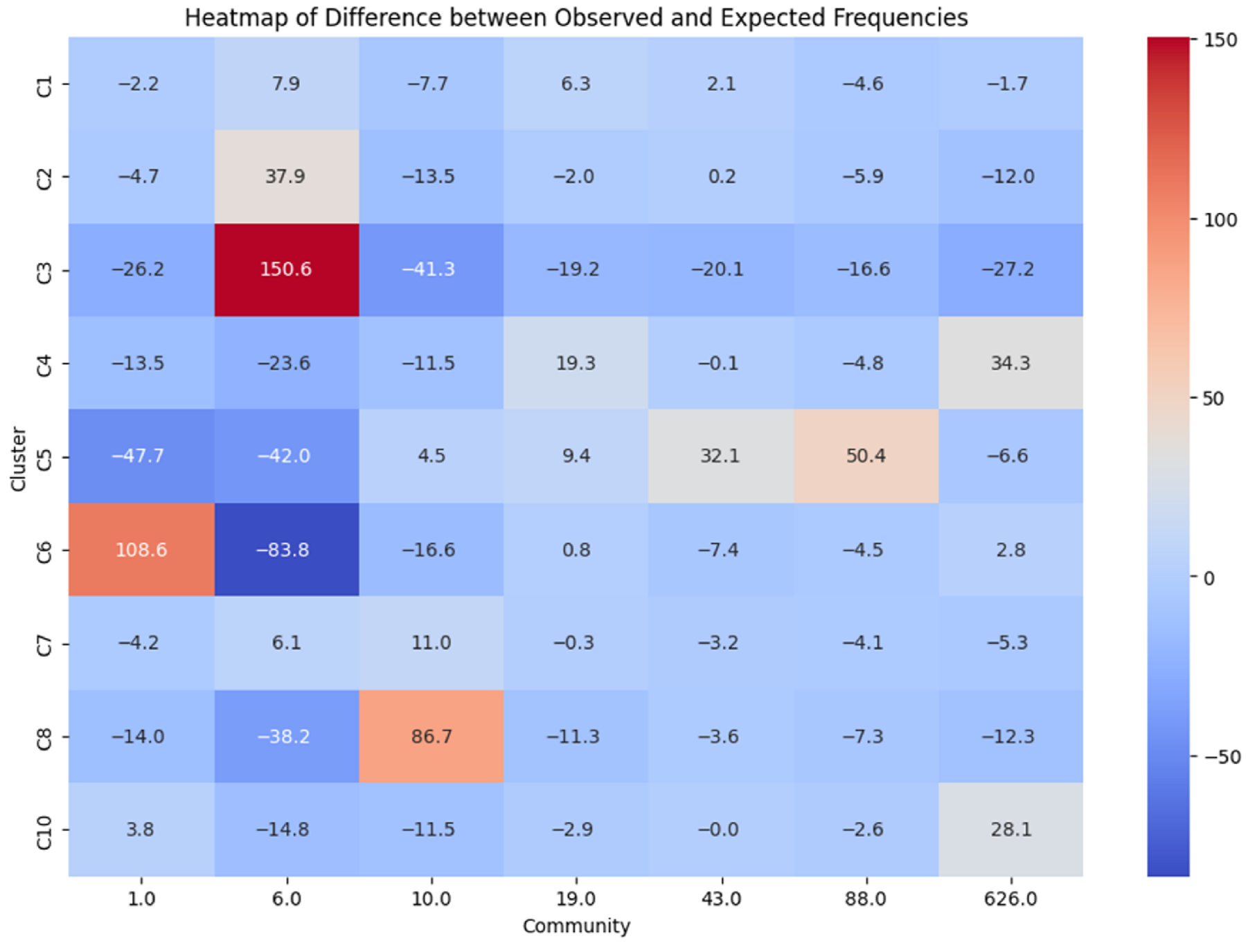

**Source(s):** Own elaboration with MatplotLib

**Figure 4.**
Associations between themes and visual narratives: differences between the frequencies observed and expected healthy eating), between cluster 6 (holistic wellbeing) and the visual narrative 1 (nature and the outdoors), and between cluster 8 (massages and beauty) and the visual narrative 10 (massages and bodywork).

Whilst it was previously difficult to identify clear visual patterns, since the focus was the images with the greatest number of likes for each cluster, a finer analysis based on advanced artificial intelligence techniques makes it possible to identify a relevant and contextually congruent association between the themes and the visual narratives classified. Even if it is a scenario of considerable narrative heterogeneity, there are visual narratives more common for certain themes and which tend to show a logical connection from the point of view of the consistency between the message written and the visual narrative.

### *4.4 The gender representation (RQ4)*

The last block of analysis has consisted of determining whether there exists a gender preference in the set of images and, more specifically, in the particular clusters that have been identified by means of the hashtag analysis. After identifying a person in 2,199 of the images in the conversation, it has been observed that 72.17% are women (1,587 images), as against 27.83% men (612 images). As this result evidences, the visual narratives in the #wellbeing field are mainly represented by women. Paying attention to community separately, the difference between genders is statistically significant ($p$-value $<0.001$) and moderately intense (Cramér's V of 0.13). Analysing the difference between the frequencies observed and those expected (Figure 5), the presence of women in the images is higher than would be expected by chance distribution in clusters 5, 7, 8 and 9 (themes of healthy nutrition,

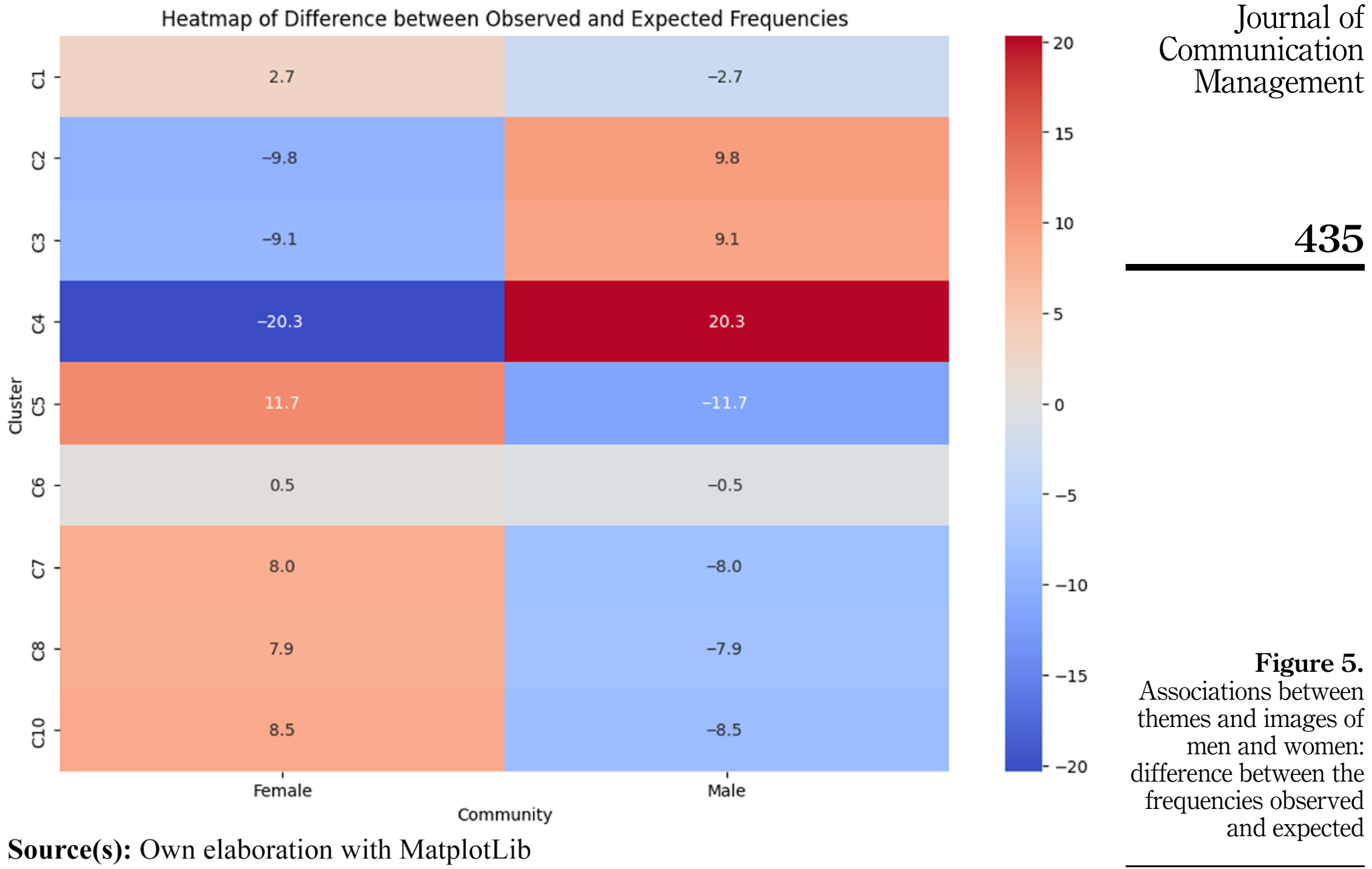

**Source(s):** Own elaboration with MatplotLib

**Figure 5.** Associations between themes and images of men and women: difference between the frequencies observed and expected

massages and beauty, holistic wellbeing and motivation), while men appear in a higher proportion than would be expected in clusters such as 2, 3 and 4 (mental health, psychological therapies and physical activity themes).

Despite the appearance of chaos hinted at by the heterogeneous, it is thus possible to identify through the themes identified gender differences. The set of visual narratives related to #wellbeing on Instagram is strongly feminised, but this feminisation is even more pronounced in some particular clusters, such as those linked to nutrition, beauty and holistic and spiritual wellbeing. On the other hand, the narratives that deal with mental health, therapies and exercise have a more balanced presence of men and women, although men do not appear in a majority in any particular cluster in terms of absolute figures.

## 5. Conclusions and discussion

The qualitative analysis carried out based on the content of the clusters and hashtags, as well as the images that accompany the posts with most likes, offers one of the first approaches carried out so far regarding the thematisation of the digital conversation about wellbeing in the social media. This study, then, goes beyond the specific sphere of previous studies which focussed on hashtags such as #goodlife (Loukianov *et al.*, 2020).

The study identifies different themes in the conversation with a high degree of complementarity, to the extent that sometimes they are indistinguishable (RQ1). Observing the dataset, there are no obvious differences, either, in terms of engagement with themes (RQ2) due to the fact that the contents tend to present a limited engagement, without it mattering in which cluster they are. It is worth mentioning that the study has a limitation in that it does not have data on the number of followers or the average engagement of users, suggesting that this could be a crucial aspect to investigate in the future. That said, the possibility of focussing on the most successful contents (in the top 100 likes per cluster, or in the top 100 comments per cluster) makes it possible to determine important differences among types of content. In fact, these differences are able to explain practically 50% of the variability of the engagement registered.

The study also identifies different narratives in the case of the images examined, that is, a heterogeneous visual narrative space in which, however, it is possible to identify certain patterns, at least when the themes of greatest interest and engagement and the most outstanding visual narratives are linked (RQ3).

In this regard, the contrast between the results of the hashtag analysis and the visual analysis make it possible to distinguish two major macro-stories (RQ1 and RQ3): on the one hand, mental health and emotional wellbeing, themes that often include hashtags about therapies, self-knowledge, spirituality or motivational phrases; on the other hand and to a lesser extent, themes and visual narratives that are related to physical activity and healthy habits, which focus on exercise and training, or on nutrition. In this way, it can be concluded that the focus of the digital conversation about #wellbeing on Instagram can be found in the mental, the psychological and the spiritual, and that both the most abundant contents and the most successful in terms of engagement are those closely linked to therapeutic narratives and to positive psychology, leaving the matter of physical activity and nutrition in a very obviously secondary position.

It is possible to argue, then, that Instagram is an interesting resource for promoting “therapeutic” actions or the improvement of people’s mental and emotional wellbeing, by means of sharing emotional narratives linked to wellbeing, which can work as therapeutic narratives. These help to give meaning to individuals’ lives and experiences of wellbeing based on stories that talk about the people themselves, which allows them to share their feelings, the intimate, with others, generating in this way alleviation when faced with certain mental and emotional concerns and anxieties.

In this regard, the problems of the individual are externalised, and a collaborative focus is created that can help prevent them becoming pathological, which in turn contributes to wellbeing. In this line, it would also be possible to link the most successful hashtags and clusters in the set analysed in the study to positive psychology, in line with the contributions regarding human wellbeing of Seligman’s PERMA model (2018). These uses can be contextualised in the terms of what Furedi (2003) has called “therapy culture”, which promotes all kinds of positive stories that help an individual’s psychology, given the idea of the expansion of emotional vulnerabilities in developed societies.

Lastly, a strongly feminised digital conversation about wellbeing has been observed (RQ4), since the presence of men and women is only balanced in some very specific clusters, such as those which are related to mental health or physical exercise. These findings open the door to future analyses regarding the feminisation of images on Instagram and their impact on the analysis of gender bias – in the sphere of wellbeing and, in general, of the social media – in relation to the use of algorithms, a research field of growing interest.

## Notes

1. https://arxiv.org/abs/2103.00020
2. https://arxiv.org/abs/0803.0476
3. https://arxiv.org/abs/2005.12872

## Further reading

## Corresponding author

Ainara Larrondo-Ureta can be contacted at: ainara.larrondo@ehu.eus